\documentclass[a4paper,12pt]{article}
\usepackage[cp1251]{inputenc}
\usepackage[english]{babel}
\usepackage{amsmath}
\usepackage{amssymb}
\usepackage{amstext}
\usepackage{euscript}
\usepackage{textcomp}
\usepackage{setspace}
\usepackage{fancyhdr}
\usepackage{graphicx}
\usepackage[pdftex]{color}
\usepackage[square,numbers,comma]{natbib}
\usepackage[pdftex]{hyperref}
\hypersetup{
             final,
             colorlinks=true,
             linkcolor=blue,
             citecolor=red
}
\begin{document}

\hspace{6cm}\textit{Kratkie Soobshcheniya po Fizike, No. 1, pp.
46-56 (2008)}

\hspace{8.7cm}[Bulletin of the Lebedev Physics Institute]

\vspace{0.2cm}
\begin{center}
{\fontfamily{ptm}\fontsize{18pt}{20pt}\selectfont{\bf GROUND STATE ENERGY\\
\vspace{0.3cm} OF CURRENT CARRIERS IN GRAPHENE}} \vspace{0.5cm}

P. V. Ratnikov\footnote{\url{ratnikov@lpi.ru}} and A. P. Silin

\vspace{0.3cm} \textit{Lebedev Physics Institute, Russian Academy of
Sciences, \\Leninskii pr. 53, Moscow, 119991 Russia}

\vspace{0.25cm}

Received October 30, 2007
\end{center}

\vspace{0.15cm}
\begin{list}{}
{\rightmargin=0cm \leftmargin=3cm}
\item
{\fontfamily{ptm}\fontsize{11pt}{8pt}\selectfont{{\bf Abstract.} The
ground state energy of current carriers in graphene considered as a
zero-gap semiconductor was calculated in the two-band approximation.
The condition of the electronic (hole) system stability in graphene
was obtained. The possibility of the zero-gap
semiconductor--semimetal transition was discussed.}}
\end{list}

\vspace{0.3cm}

It is known that thin graphite films exhibit semimetallic properties
\citep{Novoselov2004}; however, a single-atomic layer of carbon
atoms forming a regular hexagonal lattice (graphene) has such a band
structure that the energy gap is zero at six K points of the
Brillouin zone. Therefore, graphene can be considered as a
two-dimensional zero-gap semiconductor or a semimetal with zero
conduction and valence band overlap \citep{Novoselov2005}. The
former approach makes it possible to describe current carriers in
graphene within the two-band Dirac model\footnote{Dirac equation
\eqref{1} is equivalent with accuracy of the unitary transformation
of the Hamiltonian and the wave function of a pair of Weyl equations
(see book \citep{Akhiezer}, p. 79). As is known, the Weyl equation
describes the two-component neutrino in quantum electrodynamics
(QED) (see, e.g., book \citep{Schweber}). The use of the Dirac
equation as a $4\times4$ matrix equation in the two-dimensional
system is possible since the $4\times4$ matrix representation in
case of two spatial dimensions can be used equivalently with
$2\times2$ matrix representation (see book \citep{Tsvelik}, chap.
XIV). This fact allows us to extend the formalism of the QED diagram
technique to the case of the two-dimensional system of Dirac
fermions (graphene). The Weyl equation was first applied to the
problem of describing current carriers in a zero-gap semiconductor
in \citep{Nielsen}.} \citep{Volkov, Ando}
\begin{equation}\label{1}
u{\boldsymbol\alpha}\cdot\widehat{\bf p}\Psi=\varepsilon_{\bf
p}\Psi,
\end{equation}
where
${\boldsymbol\alpha}=\left(
\begin{matrix}
0&{\boldsymbol\sigma}\\ 
{\boldsymbol\sigma}&0
\end{matrix}
\right)$
are Dirac $\alpha$-matrices, $\widehat{\bf
p}=-i\hbar{\boldsymbol\nabla}$ is the two-dimensional momentum
operator (hereafter $\hbar=1$), $u=\frac{3}{2}\gamma
a_0=9.84\cdot10^7$ cm/s\, is the quantity similar to the Kane matrix
element of the interband transition rate, $\gamma\simeq3$ eV is the
band parameter, and $a_0=1,44 \ \text{\AA}$ is the interatomic
distance in the graphene lattice \citep{Falkovsky}. In the vicinity
of K points of the Brillouin zone, the dispersion relation of
current carriers is linear, $\varepsilon_p=\pm up$ ($+$ and $-$
signs correspond to electrons and holes, respectively).

For a two-dimensional electron (hole) gas arising during electron
doping of a zero-gap semiconductor \citep{Novoselov2005}, the
ground state energy per one particle is the sum of three terms
\begin{equation}
E_{gs}=E_{kin}+E_{exch}+E_{corr},
\end{equation}
where $E_{kin}=\frac{2}{3}up_F$ is the average kinetic energy,
$p_F=\sqrt{\frac{2\pi n_{2D}}{\nu}}$\, is the Fermi momentum,
$n_{2D}$ is the two dimensional particle concentration, $\nu$ is the
degeneration multiplicity\footnote{In the general case for two spin
components, the degeneration multiplicity is $\nu=\nu_{e,h}$. It
will be shown below that the spin-unpolarized state is more
energetically favorable than spin-polarized one, for which
$\nu\rightarrow\nu_{e,h}/2$.}. If the Fermi level $E_F$ lies above
$E=0$, the system contains only electrons as current carriers in the
conduction band with the number of valleys $\nu_e=2$; if $E_F<0$,
the system contains only holes as current carriers with $\nu_h=2$.
The Fermi level position can be varied by applying an electric field
\citep{Novoselov2005}. We can see that both cases in the Dirac model
are equivalent. In follows, for definiteness, we shall consider the
case of electrons.

The exchange energy is given by the first-order exchange diagram (\hyperlink{fig1}{Fig. 1})
\begin{equation}\label{3}
E_{exch}=-\frac{\nu}{2n_{2D}}\int\frac{d^2{\bf p}d\varepsilon}{(2\pi)^3}
\frac{d^2{\bf k}d\omega}{(2\pi)^3}Sp\left\{\gamma^\mu
G\left({\bf p}, \varepsilon\right)\gamma^\nu
G\left({\bf k},
\omega\right)\right\}D_{\mu\nu}\left({\bf p-k},
\varepsilon-\omega\right),
\end{equation}
where the photon propagator $D_{\mu\nu}\left({\bf p-k},
\varepsilon-\omega\right)\approx
V({\bf p-k})\delta_{\mu4}\delta_{\nu4}$
(we neglect the photon pole at $\omega=\pm
c\left|{\bf p-k}\right|$, whose contribution to the integral in frequencies
$\varepsilon$ and $\omega$ is of the order of $\left(u/c\right)^2\sim10^{-5}$ in comparison
with the contribution of Green's function poles), $V({\bf q})=\frac{2\pi
e^2}{\kappa_{eff}|{\bf q}|}$ is the Coulomb law in the two-dimensional case, and
$\kappa_{eff}$ is the effective permittivity of graphene. The free-electron Green's function at
$\Delta=0$ is \citep{Pechenik}
\begin{equation}
G({\bf p},
\varepsilon)=-\frac{u\widehat{p}}{u^2{\bf p}^2-\varepsilon^2-i0}+2\pi
i\delta\left(u^2{\bf p}^2-\varepsilon^2\right)N_pu\widehat{p},
\end{equation}
where $\widehat{p}=p_\beta\gamma^\beta\, (\beta=0, 1, 2)$ is the
convolution with Dirac matrices $\gamma^k=-i\gamma^0\alpha^k$ for
$k=1, 2$ and
$\gamma^0=\left(\begin{matrix}I&0\\0&-I\end{matrix}\right), I$ is
the $2\times2$ unit matrix, $N_p=\theta(|{\bf
p}|-p_F)\theta(\varepsilon)$; $\theta(x)=\begin{cases}1, \ x>0, \\
0, \ x<0.\end{cases}$

Expression \eqref{3} is transformed to the form (see the
\hyperlink{App}{Appendix})
\begin{equation}\label{5}
E_{exch}=-\frac{\alpha^*I_1}{\sqrt{2\pi}}u\left(\frac{n_{2D}}{\nu}\right)^{1/2},
\end{equation}
\begin{center}
\hypertarget{fig1}{}
\includegraphics[width=5cm]{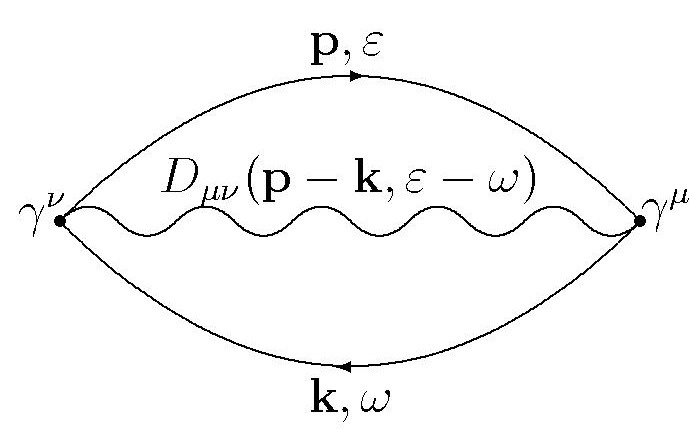}
\vspace{0.25cm}

Fig. 1. \emph{First-order exchange diagram.}
\end{center}
where $I_1=\int\limits_0^1dx\int\limits_0^1dy\int\limits_0^{2\pi}d\chi\frac{(1+\cos\chi)xy}{\sqrt{x^2+y^2-2xy\cos\chi}}=
\frac{8}{3}\left(G+\frac{1}{2}\right), \ G=0,915965\ldots$ is the Catalan constant \citep{Gradshtein},
$\alpha^*=\frac{e^2}{\kappa_{eff} u}$ is the analogue of the fine structure constant.

The correlation energy is given by the formula \citep{Keldysh}
\begin{equation}\label{6}
E_{corr}=\frac{1}{2n_{2D}}\int\frac{d^2{\bf k}d\omega}{(2\pi)^3}\int\limits_0^1\frac{d\lambda}{\lambda}
\left[\frac{-\lambda\nu
V\left({\bf k}\right)\Pi_{44}\left({\bf k},
i\omega\right)}{1-\lambda\nu
V\left({\bf k}\right)\Pi_{44}\left({\bf k},
i\omega\right)}+\lambda\nu
V\left({\bf k}\right)\Pi^{(0)}_{44}\left({\bf k},
i\omega\right)\right].
\end{equation}
The total polarization operator is written as
\begin{equation}\label{7}
\Pi_{44}\left({\bf k},
i\omega\right)=\Pi^{(0)}_{44}\left({\bf k},
i\omega\right)+\Pi^{(1)}_{44}\left({\bf k},
i\omega\right)+\ldots,
\end{equation}
which corresponds to the sum of the diagrams
\begin{center}
\includegraphics[width=15cm]{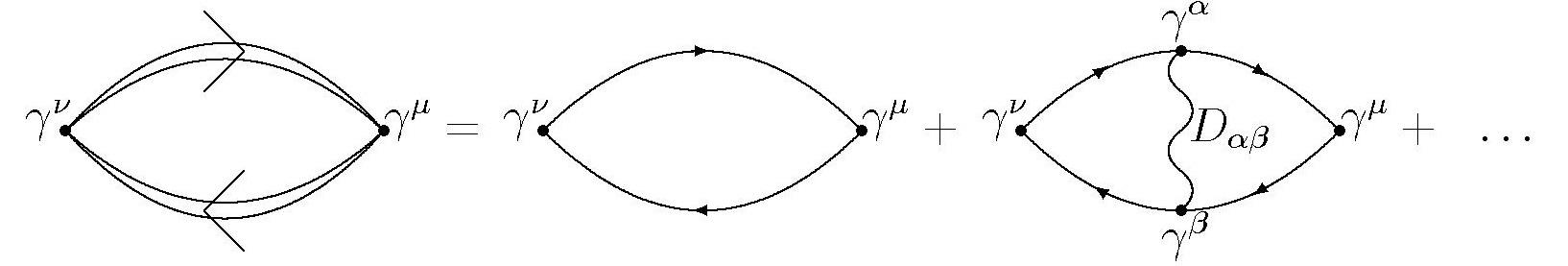}
\end{center}

The polarization operator $\Pi^{(0)}_{44}({\bf k}, i\omega)$ in the
lowest order in the interaction for the two-dimensional case is
given by \citep{Pechenik}
\begin{equation}\label{8}
\Pi^{(0)}_{44}\left({\bf k},
i\omega\right)=16\int\frac{d^2{\bf p}}{(2\pi)^2}\frac{\theta\left(\left|{\bf p}\right|-p_F\right)}{2\varepsilon_{\bf p}}
\frac{\left({\bf k}\cdot{\bf p}\right)^2-\frac{\left|{\bf k}\right|^2\varepsilon^2_{\bf p}}{u^2}}
{\left(\left|{\bf k}\right|^2+\frac{\omega^2}{u^2}\right)^2-4\left({\bf k}\cdot{\bf p}-\frac{i\varepsilon_{\bf p}\omega}{u^2}\right)^2}.
\end{equation}
We calculate the correlation energy using a method similar in many
respects to the known Nozieres--Pines method \citep{Nozieres,
Combescot} which is applied to calculate the electron gas and
electron-hole liquid using asymptotics of the dimensionless
polarization operator \citep{Andryushin}
\begin{equation}\label{9}
\widetilde{\Pi}^{(0)}_{44}(r,
\theta)=\begin{cases}-\frac{\sin\theta}{24\pi r^2}, \ r\gg1,
\\ -\frac{1-|\cos\theta|}{16\pi}, \ r\ll1,\end{cases}
\end{equation}
where dimensionless variables $r=\sqrt{\omega^2+u^2k^2}/up_F$ and $\sin\theta=uk/\sqrt{\omega^2+u^2k^2}$ are introduced.

To determine the smallness parameter of expansion \eqref{7}, we estimate the first-order correction
$\Pi^{(1)}_{44}$ in the interaction to the polarization operator \eqref{8}
$$
\Pi^{(1)}_{44}\left({\bf k},
i\omega\right)=-\int\frac{d^2{\bf p}d\varepsilon}{(2\pi)^3}\frac{d^2{\bf q}d\Omega}{(2\pi)^3}
Sp\left\{\gamma^\mu G\left({\bf p+q},
\varepsilon+\Omega\right)\gamma^\alpha G\left({\bf p},
\varepsilon\right)\cdot\right.
$$
\begin{equation}\label{10}
\left.\cdot\gamma^\nu G\left({\bf p-k},
\varepsilon-\omega\right)\gamma^\beta
G\left({\bf p+q-k},
\varepsilon+\Omega-\omega\right)\right\}\times
V({\bf q})\delta_{\alpha4}\delta_{\beta4}\approx
\end{equation}
$$
\approx-\frac{1}{16\pi
up^2_F}\int\frac{d^2{\bf p}d^2{\bf q}}{(2\pi)^4}\frac{2\pi\alpha^*}{\left|{\bf q}\right|}
\theta\left(p_F-\left|{\bf p}\right|\right)\theta\left(p_F-\left|{\bf q}\right|\right)=
-\frac{\alpha^*}{(8\pi)^2}\frac{p_F}{u}.
$$
Taking into account that the main contribution to
$\Pi^{(1)}_{44}\left({\bf k}, i\omega\right)$ is made by small
transferred momenta due to $V({\bf q})$ \citep{Andryushin},
\eqref{10} should be compared with the asymptotics
$\Pi^{(0)}_{44}({\bf k}, i\omega)$ at small $\left|{\bf k}\right|$,
from which we obtain that \eqref{10} is small in the parameter
$\frac{\alpha^*}{4\pi}\ll1$, which is simultaneously the condition
of the applicability of the random-phase approximation.

Substitution of the polarization operator asymptotics at $r\gg1$ into \eqref{6} yields the contribution of large momenta
\begin{equation}\label{11}
E^\infty_{corr}=-\frac{up^3_F}{3(2\pi)^2n_{2D}}[(1+g_1)\cdot\ln(1+g_1)-g_1],
\end{equation}
where $g_1=\frac{\alpha^*\nu}{12}\ll1$ at $\nu=2$ and
$\alpha^*\lesssim1$; expanding \eqref{11}, we obtain
\begin{equation}\label{12}
E^\infty_{corr}=-\frac{\alpha^{*2}\nu^{1/2}}{864\sqrt{2\pi}}un^{1/2}_{2D},
\end{equation}
which coincides with the contribution of the second-order ring diagram (\hyperlink{fig2-3}{Fig. 2})
$$
E^{(2)}_{corr}=-\frac{i\nu^2}{4n_{2D}}\int\frac{d^2{\bf p}d^2{\bf q}d^2{\bf k}
d\varepsilon d\Omega d\omega}{(2\pi)^9}Sp\left\{\gamma^\alpha
G\left({\bf p}, \varepsilon\right)\gamma^\beta
G\left({\bf p-k},
\varepsilon-\omega\right)\right\}\times
$$
\begin{equation}\label{13}
\times Sp\left\{\gamma^\mu G\left({\bf q},
\Omega\right)\gamma^\nu
G\left({\bf q+k},
\Omega+\omega\right)\right\}D_{\alpha\mu}\left({\bf k},
\omega\right)D_{\beta\nu}\left({\bf k},
\omega\right)\approx
\end{equation}
$$
\hspace{-4cm}\approx-\frac{1}{4n_{2D}}\int\frac{d^2{\bf k}d\omega}{(2\pi)^3}\left(\nu\Pi^{(0)}_{44}
\left({\bf k},
i\omega\right)V\left({\bf k}\right)\right)^2.
$$
Thus, according to the Nozieres--Pines method \citep{Nozieres,
Combescot}, when calculating the correlation energy at large
transferred momenta by formula \eqref{6}, with an accuracy of the
terms of the order of $g^2_1$, the analysis can be restricted to the
second-order of the perturbation theory. Apart from the second-order
ring diagram (\hyperlink{fig2-3}{Fig. 2}), let us also consider the
second-order exchange diagram (\hyperlink{fig2-3}{Fig. 3})
\begin{equation*}
\widetilde{E}^{(2)}_{corr}=-\frac{i\nu}{4n_{2D}}\int\frac{d^2{\bf
p}d^2{\bf q}d^2{\bf k} d\varepsilon d\Omega
d\omega}{(2\pi)^9}Sp\left\{\gamma^\alpha G\left({\bf p},
\varepsilon\right)\gamma^\beta G\left({\bf k-p},
\omega-\varepsilon\right)\gamma^\mu\cdot\right.
\end{equation*}
\begin{equation}\label{14}
\cdot\left.
G\left({\bf p-q-k},
\varepsilon-\Omega-\omega\right)\gamma^\nu
G\left({\bf p-q},
\varepsilon-\Omega\right)\right\}D_{\alpha\mu}\left({\bf q},
\Omega\right)D_{\beta\nu}\left({\bf k}, \omega\right).
\end{equation}
Evaluation of integral \eqref{14} is very laborious; however, estimations show that, as in the nonrelativistic case, its contribution is positive and is
smaller in magnitude by a factor of $\frac{1}{2\nu}$ than \eqref{13} \citep{Pines}. Finally, for the contribution of large transferred momenta to the correlation energy, we obtain
\begin{equation}\label{15}
E^\infty_{corr}=-\frac{\alpha^{*2}\nu^{1/2}}{864\sqrt{2\pi}}\left(1-\frac{1}{2\nu}\right)un^{1/2}_{2D}.
\end{equation}
Substitution of the polarization operator asymptotics at $r\ll1$
from \eqref{9} into \eqref{6} yields the contribution of small
transferred momenta
\begin{center}
\hypertarget{fig2-3}{}
\includegraphics[width=15cm]{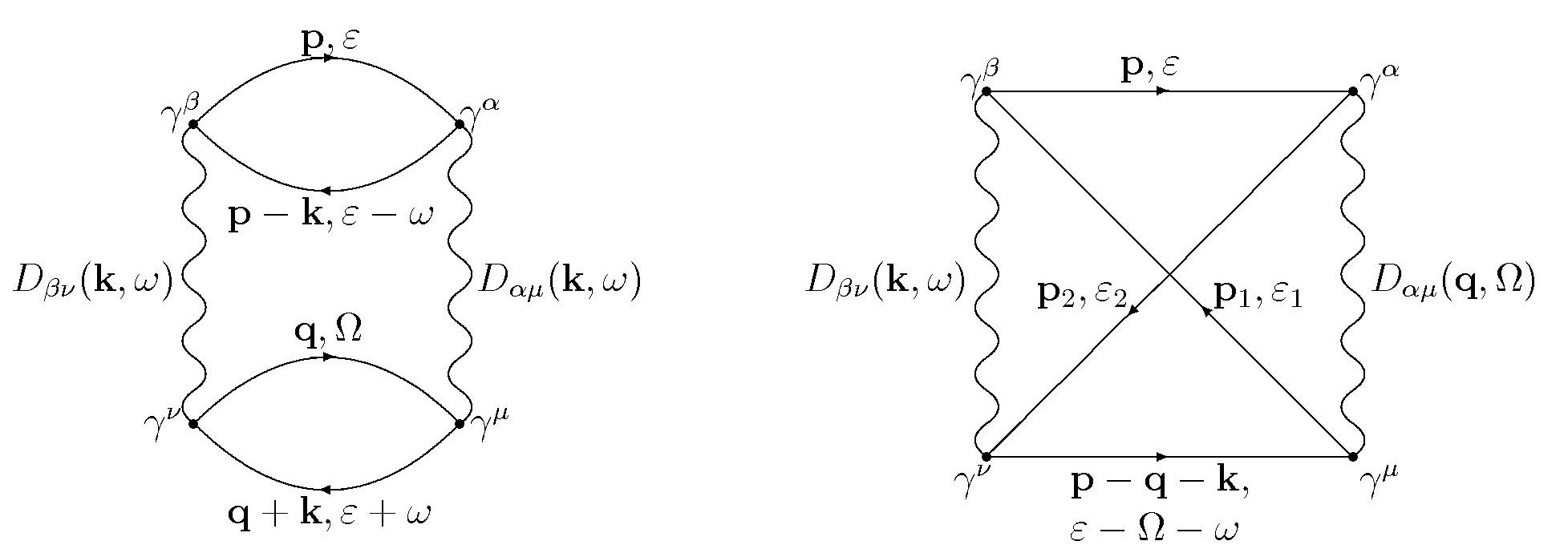}
\end{center}

Fig. 2. \emph{Second-order ring diagram.}\hspace{2cm}Fig. 3.
\emph{Second-order exchange diagram.}

\hspace{8.2cm}\emph{The following notations are introduced:}

\hspace{7.5cm}${\bf p}_1={\bf p-k}, \
\varepsilon_1=\varepsilon-\omega;\ {\bf p}_2={\bf p-q}, \
\varepsilon_2=\varepsilon-\Omega.$
\begin{equation*}
E^0_{corr}=-\frac{up^3_Fg^2_2}{3(2\pi)^2n_{2D}}\int\limits_0^{\pi/2}d\theta(1-\cos\theta)^2\cdot\left[1-
g_2\frac{1-\cos\theta}{\sin\theta}\cdot\ln\left(1+\frac{\sin\theta}{g_2(1-\cos\theta)}\right)+\right.
\end{equation*}
\begin{equation}\label{16}
\left.+\frac{\sin\theta}{g_2(1-\cos\theta)}-\frac{\sin^2\theta}{g^2_2(1-\cos\theta)^2}\cdot\ln
\left(1+g_2\frac{1-\cos\theta}{\sin\theta}\right)\right],
\end{equation}
where $g_2=\frac{\alpha^*\nu}{8}\ll1$, and expression \eqref{16} can be simplified,
\begin{equation}\label{17}
E^0_{corr}=-\frac{\alpha^{*2}\nu^{1/2}}{128\sqrt{2\pi}}\left(\frac{3\pi}{4}-2\right)un^{1/2}_{2D}.
\end{equation}
We can see from \eqref{15} and \eqref{17} that the order of smallness of the correlation energy and exchange energy
\eqref{5} is $\alpha^{*2}$ and $\alpha^*$, respectively; thus, we obtain the ground state energy of the electron gas as a power series of $\alpha^*$, which
we cut off after the terms of the order $\alpha^{*2}$,
\begin{equation}\label{18}
E_{gs}=\frac{2\sqrt{2\pi}}{3}u\left(\frac{n_{2D}}{\nu}\right)^{1/2}-\frac{\alpha^*I_1}{\sqrt{2\pi}}
u\left(\frac{n_{2D}}{\nu}\right)^{1/2}-\frac{\alpha^{*2}}{64\sqrt{2\pi}}\left(\frac{3\pi}{8}-
\frac{25}{27}-\frac{1}{27\nu}\right)u\left(\nu n_{2D}\right)^{1/2}.
\end{equation}
It can be seen from \eqref{18} that the spin-unpolarized state with
the degeneracy multiplicity $\nu=\nu_{e,h}$ is more energetically
favorable than the spin-polarized state with the degeneracy
multiplicity $\nu=\nu_{e,h}/2$. The main contribution is given by
the kinetic energy; therefore, $E_{gs}>0$. The condition $E_{gs}>0$
in the case of $\Delta=0$ means that graphene as a zero-gap
semiconductor is $\textit{stable}$ (at $E_{gs}<0$, it would be
favorable to generate electron--hole pairs). The transition to the
semimetallic state, i.e., spontaneous electron--hole pair
generation, occurs at certain $\alpha^*_0$ such that $E_{gs}<0$ at
$\alpha^*>\alpha^*_0$. Equation $E_{gs}=0$ upon the substitution of
$\nu=2$ yields
\begin{equation}\label{19}
\alpha^*_0=\sqrt{A^2+B}-A=1.1044,
\end{equation}
where $A=\frac{64I_1}{3\pi-202/27}$\, and
$B=\frac{512\pi}{9\pi-202/9}$.

Let us now consider the effect of some parameters on the band structure of graphene in more detail.

$\textit{Effective multivalley structure.}$ A structure
(superlattice) containing $N$ graphene layers, in the absence of
transitions  between layers, effectively contains
$\widetilde{\nu}=\nu\cdot N$ electron (hole) valleys with the number
of electrons (holes) $\overline{N}_{e(h)}=N_{e(h)}\cdot N$, where
$N_{e(h)}$ is the number of electrons (holes) in each graphene
layer. Let graphene layers be interfaced by a wide-gap semiconductor
(insulator).

In the general case the Coulomb law for periodic structures is given by the expression \citep{Andryushin}
\begin{equation}\label{20}
\overline{V}\left({\bf q}, w\right)=\frac{2\pi e^2}{\kappa\left|{\bf
q}\right|} \frac{\sinh{\left(\left|{\bf
q}\right|d\right)}}{\cosh{\left(\left|{\bf
q}\right|d\right)}-\cos{w}},
\end{equation}
where $0\leq w\leq 2\pi$; however, in the case of large transferred momenta such that $\left|{\bf q}\right|d\gg1$, the second fraction in \eqref{20} tends to unity.
Formula \eqref{11} for $E^\infty_{corr}$ remains valid with an accuracy of the substitution $g_1\rightarrow\widetilde{g}_1=\frac{\alpha^*\widetilde{\nu}}{12}\gg1$ and
$n_{2D}\rightarrow\overline{n}_{2D}=\overline{N}_{e(h)}/S=N\cdot
n_{2D}$ ($S$ is the area of layers),
\begin{equation}\label{21}
E^\infty_{corr}=-\frac{\alpha^*}{36\sqrt{2\pi}}u\left(\frac{n_{2D}}{\nu}\right)^{1/2}\cdot\ln
\left(\frac{\alpha^*\nu}{12}N\right).
\end{equation}
At small transferred momenta, $\left|{\bf q}\right|d\ll1$, the
Coulomb law \eqref{20} is not singular at $\left|{\bf q}\right|=0$.
In this limit, $E^0_{corr}$ is small in comparison with
$E^\infty_{corr}$ due to appearance of $\ln N$; therefore, it can be
omitted. The ground state energy is given by
\begin{equation}\label{22}
E_{gs}=\left[\frac{2\sqrt{2\pi}}{3}-\frac{\alpha^*}{\sqrt{2\pi}}\left(I_1+\frac{1}{36}
\ln\left(\frac{\alpha^*\nu}{12}N\right)\right)\right]\cdot
u\left(\frac{n_{2D}}{\nu}\right)^{1/2}.
\end{equation}

$\textit{Electric field effect.}$ As shown above, application of an electric field can produce a nonzero electron (or holes when the electric field direction is changed)
density in graphene on substrate \citep{Novoselov2005}. In this case, the two-dimensional current-currier concentration is proportional to the gate voltage $V_g$ \citep{Novoselov2005}
\begin{equation}\label{23}
n_{2D}=\frac{\epsilon}{4\pi el}V_g,
\end{equation}
where $\epsilon$ is the substrate permittivity.

According to the calculation \citep{Aoki}, in graphene containing
several layers, the energy gap opens at K points of the Brillouin
zone under a sufficiently strong electric field. This means that the
system transits from a semimetallic state (studied in
\citep{Novoselov2004}) to the semiconductor state. Let us find out
whether or not a similar phenomenon exists in single-layer graphene.
Assume that application of a rather weak electric field results in
opening the gap, $\Delta\neq0$, small in comparison with the Fermi
energy $E_F=up_F$: $\Delta\ll
E_F=\frac{u}{2}\sqrt{\frac{\epsilon}{el}V_g}$; in this case, the
correction to the kinetic energy is $\delta E_{kin}=\Delta^2/E_F$.
According to (\hyperlink{A1}{A1}), the correction to the exchange
energy is given by
\begin{equation}\label{24}
\delta E_{exch}=-\frac{\alpha^*I_2}{2\pi}\frac{\Delta^2}{E_F},
\end{equation}
where $I_2=\int\limits_0^1dx\int\limits_0^1dy\int\limits_0^{2\pi}d\chi
\frac{1}{\sqrt{x^2+y^2-2xy\cos\chi}}=16G$. Let us express the correction to the correlation energy in terms of the correction to the polarization operator $\overline{\Pi}^{(0)}_{44}\left({\bf k},
i\omega\right)=\Pi^{(0)}_{44}\left({\bf k},
i\omega\right)+\delta\Pi^{(0)}_{44}\left({\bf k},
i\omega\right)$, omitting the terms giving a zero contribution to $\delta
E_{corr}$ (odd-numbered in $\omega$)
\begin{equation}\label{25}
\delta\Pi^{(0)}_{44}\left({\bf k},
i\omega\right)=-\frac{3}{4\pi}\frac{u^2\left|{\bf
k}\right|^4\Delta^2} {\omega^3\left(u^2\left|{\bf
k}\right|^2+\omega^2\right)}
\arctan{\left(\frac{2up_F\omega}{u^2\left|{\bf
k}\right|^2+\omega^2}\right)},
\end{equation}
\begin{equation}\label{26}
\delta
E_{corr}=-\frac{\nu^2}{2n_{2D}}\int\frac{d^2{\bf k}d\omega}{(2\pi)^3}
\frac{V^2\left({\bf k}\right)\Pi^{(0)}_{44}\left({\bf k},
i\omega\right)}{1-\nu
V\left({\bf k}\right)\Pi^{(0)}_{44}\left({\bf k},
i\omega\right)}\delta\Pi^{(0)}_{44}\left({\bf k},
i\omega\right).
\end{equation}
Changing to dimensionless variables $r, \theta$ and substituting the asymptotics $\widetilde{\Pi}^{(0)}_{44}(r, \theta)$ at $r\gg1$ and $r\ll1$, noticing that the main contribution to
$\delta E_{corr}$ given by small transferred momenta due to $V^2({\bf k})$ in \eqref{26}, we obtain
\begin{equation}\label{27}
\delta
E_{corr}=\frac{9\pi\alpha^{*2}\nu}{256}\ln\left(1+\frac{8}{\alpha^*\nu}\right)\cdot\frac{\Delta^2}{E_F}.
\end{equation}
The correction to the ground state energy contains the additional small factor
\begin{equation}\label{28}
\delta E_{gs}=\delta E_{kin}+\delta E_{exch}+\delta
E_{corr}=\left[1-\frac{8G\alpha^*}{\pi}+\frac{9\pi\alpha^{*2}\nu}{256}\ln\left(1+\frac{8}{\alpha^*\nu}\right)\right]
\frac{\Delta^2}{E_F}\approx0.1202\frac{\Delta^2}{E_F};
\end{equation}
therefore,even in the presence of a gap (small in comparison with
$E_F$), expression \eqref{18} is correct; moreover, this correction
is positive, which suggests that the zero-gap semiconductor phase is
$\textit{stable}$ relative to the transition to nonzero-gap
semiconductor phase in an external electric field.

$\textit{Transition to the semimatallic phase.}$ For graphene, the
parameter $\alpha^*$ can efficiently vary due to image forces, i.e.,
the variation in the effective permittivity $\kappa_{eff}$ of
graphene depending on its environment (insulator or vacuum); at the
substrate  thickness $l\gg n^{-1/2}_{2D}$, it is given by
$\kappa_{eff}=\frac{\epsilon+\epsilon^\prime}{2}$, where
$\epsilon^\prime$ is the permittivity of a medium above graphene
\citep{Keldysh1}. For a $SiO_2$ substrate, $\kappa_{eff}=5$ and
$\alpha^*\approx0,44$; for a $SiC$ substrate, $\kappa_{eff}=3$ and
$\alpha^*\approx0,73$ \citep{Iyengar}. Let us estimate the valence
and conduction band overlap $\delta E$ in the semimetallic state
(\hyperlink{fig4}{Fig. 4}). Let initially $E_F>0$ and $\Delta=0$;
and after the transition, the conduction band was lowered with
respect to the level $E=0$ by $\delta E/2$ and the valence band rose
by $\delta E/2$; then the number $\Delta N_e$ of electrons
transferred to the valence band is $\frac{\nu}{2\pi} p^2_1S$; the
new Fermi momentum is $p_2: \
N^\prime_e=\frac{\nu}{2\pi}p^2_2S=N_e+\Delta N_e$, where
$N_e=\frac{\nu}{2\pi}p^2_FS$ ($p_F$ is the Fermi momentum before the
transition). We can estimate $up_1\simeq\delta E/2$. The average
kinetic energy of electrons is $E^e_{kin}=\frac{2}{3}up_2$; for
holes, $E^h_{kin}=\frac{2}{3}up_1$ (the number of holes is equal to
the number of transferred electrons, $N_h=\Delta N_e$; therefore,
their Fermi momentum is equal to $p_1$). The ground state energy in
the semimetallic phase is given by
\begin{equation}\label{29}
E^\prime_{gs}=E^e_{kin}+E^h_{kin}+E_{exch}+E_{corr},
\end{equation}
\begin{center}
\hypertarget{fig4}{}
\includegraphics[width=15cm]{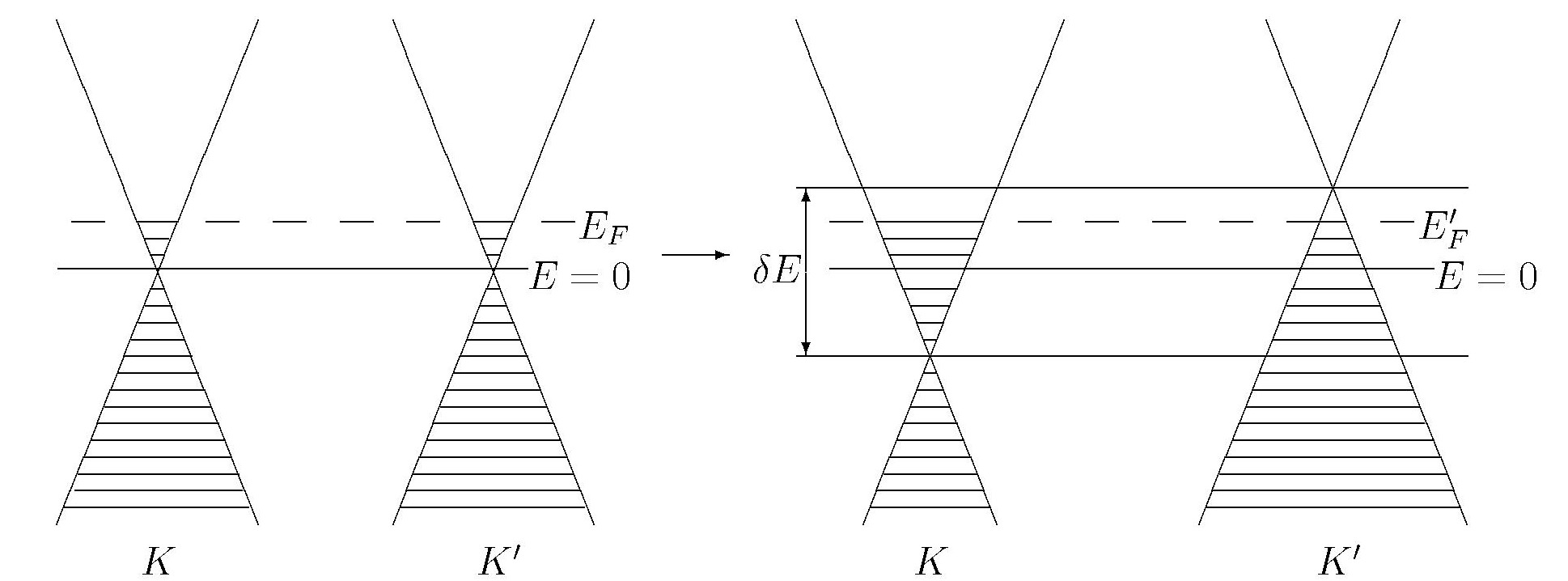}
\end{center}
Fig. 4. \emph{Transition to the semimetallic state of graphene
depending on a substrate material: the conduction band is overlapped
with the valence band at all three pairs of neighboring points $K$
and $K^\prime$ of the Brillouin zone.}

\vspace{0.25cm}\hspace{-0.6cm}where $E_{corr}$ is calculated by the
polarization operator
$\Pi^{(0)}_{44}=\Pi^{(0)e}_{44}+\Pi^{(0)h}_{44}$; \ $E_{exch}$ \ is
set equal to \eqref{5} if $N_h\simeq N_e$ so that $E^h_{exch}\simeq
E^e_{exch}=E_{exch}$. Neglecting $E_{corr}$, on the assumption that
$E^\prime_{gs}=0$, we obtain
\begin{equation}\label{30}
\delta E\simeq\left(b-\frac{1}{b}\right)up_F,
\end{equation}
where $b=\frac{|E_{exch}|}{E_{kin}}=\frac{3\alpha^*I_1}{4\pi}=\frac{2\alpha^*}{\pi}\left(G+\frac{1}{2}\right)$. The condition $b>1$ means that the zero-gap semiconductor phase is
$\textit{unstable}$ ($E_{gs}<0$); in this case, $\delta E>0$, i.e., the transition to the semimetallic state, the transition from the spin-unpolarized to the spin-polarized occurs.

This study was supported by the "Dynasty"\, foundation for noncommercial programs.

\begin{center}
\hypertarget{App}{\bf Appendix}
\end{center}

Using the Green's function from \citep{Pechenik}, we obtain the
expression for the exchange energy at arbitrary energy gap $\Delta$
\begin{equation*}\hypertarget{A1}
E_{exch}=-\frac{\nu}{2n_{2D}}\int\frac{d^2{\bf p}d^2{\bf
k}}{(2\pi)^4} \frac{\Delta^2+u^2{\bf p}\cdot{\bf k}+\varepsilon_{\bf
p}\varepsilon_{\bf k}}{\varepsilon_{\bf p}\varepsilon_{\bf k}}
\frac{2\pi e^2}{\kappa_{eff}\left|{\bf p-k}\right|}
\theta\left(p_F-\left|{\bf
p}\right|\right)\theta\left(p_F-\left|{\bf k}\right|\right). \eqno
(\text{A1})\hypertarget{A1}{}
\end{equation*}
From (\hyperlink{A1}{A1}), in the nonrelativistic limit, $\Delta\gg
up_F, \varepsilon_{\bf p}\varepsilon_{\bf k}\approx\Delta^2\gg
u^2\left|{\bf p}\cdot{\bf k}\right|$, we obtain
\begin{equation*}\hypertarget{A2}
E^{nonrel}_{exch}=-\frac{\nu}{n_{2D}}\int\frac{d^2{\bf p}d^2{\bf
k}}{(2\pi)^4} \frac{2\pi e^2}{\kappa_{eff}\left|{\bf p-k}\right|}
\theta\left(p_F-\left|{\bf
p}\right|\right)\theta\left(p_F-\left|{\bf
k}\right|\right),\eqno(\text{A2})
\end{equation*}
which coincides with the known expression (see, e.g.,
\citep{Pines}). In the ultrarelativistic limit, $\Delta\ll up_F,
\varepsilon_{\bf p}\varepsilon_{\bf k}\approx u^2\left|{\bf
p}\right|\cdot\left|{\bf k}\right|\gg\Delta^2$, we obtain
\begin{equation*}\hypertarget{A3}
E^{ultrarel}_{exch}=-\frac{\nu}{2n_{2D}}\int\frac{d^2{\bf p}d^2{\bf
k}}{(2\pi)^4} \left(1+\frac{{\bf p}\cdot{\bf k}}{\left|{\bf
p}\right|\cdot\left|{\bf k}\right|}\right) \frac{2\pi
e^2}{\kappa_{eff}\left|{\bf p-k}\right|} \theta\left(p_F-\left|{\bf
p}\right|\right)\theta\left(p_F-\left|{\bf k}\right|\right).
\eqno(\text{A3})
\end{equation*}
Dedimensionalizing the integrand in (\hyperlink{A3}{A3}) and integrating over momenta, we obtain the answer in the form of \eqref{5}.
Expression (\hyperlink{A3}{A3})
is equivalent to formula (7) in \citep{Dharma}.


\newpage
\begin{thebibliography}{99}
\bibitem[()]{Novoselov2004}
K. S. Novoselov, A. K. Geim, S. V. Morozov et al. Science, {\bf
306}, 666 (2004).
\bibitem[()]{Novoselov2005}
K. S. Novoselov, A. K. Geim, S. V. Morozov et al. Nature, {\bf 438},
197 (2005).
\bibitem[()]{Akhiezer}A. I. Akhiezer and V. B. Berestetskii,
\textsl{Quantum Electrodynamics} (Nauka, Moscow, 1969) [in
Russian].
\bibitem[()]{Schweber}S. S. Schweber, \textsl{Introduction to
Relativistic Quantum Field Theory} (Halper and Row, New York,
1961).
\bibitem[()]{Tsvelik}A. M. Tsvelik,
\textsl{Quantum Field Theory in Condensed Matter Physics} (Cambridge
University Press, 2003).
\bibitem[()]{Nielsen}H. Nielsen and M. Ninomiya, Phys. Lett. {\bf
130B}, 389 (1983).
\bibitem[()]{Volkov}B. A. Volkov, B. G. Idlis, and M. Sh. Usmanov,
\href{http://ufn.ru/ufn95/ufn95_7/Russian/r957d.pdf}
{Usp. Fiz. Nauk {\bf 65}, 799 (1995)} [Phys. Usp.].
\bibitem[()]{Ando}T. Ando, J. Phys. Soc. Jpn. {\bf 74}, 777
(2005).
\bibitem[()]{Falkovsky}L. A. Falkovsky and A. A. Varlamov,
\href{http://arxiv.org/PS_cache/cond-mat/pdf/0606/0606800v1.pdf}{Cond-mat/0606800.}
\bibitem[()]{Pechenik}L. E. Pechenik and A. P. Silin, Kratk. Soobshch. Fiz. No. 5-6, 72 (1996)
[Bull. Lebedev Phys. Inst.].
\bibitem[()]{Gradshtein}I. S. Gradshtein and I. M. Ryzhik, \textsl{Tables of Integrals, Sums, Series,
and Products} (Fizmatlit, Moscow, 1963) [in Russian].
\bibitem[()]{Keldysh}L. V. Keldysh, Contemp. Phys. {\bf 27} (5), 395 (1986).
\bibitem[()]{Nozieres}P. Nozieres and D. Pines, \href{http://prola.aps.org/pdf/PR/v111/i2/p442_1}
{Phys. Rev. {\bf 111}, 442 (1958).}
\bibitem[()]{Combescot}M. Combescot and P. Nozieres, J. Phys. C {\bf
5}, 2369 (1972).
\bibitem[()]{Andryushin}E. A. Andryushin, L. E. Pechenik, and A. P. Silin, Kratk. Soobshch. Fiz. No. 7-8, 68 (1996)
[Bull. Lebedev Phys. Inst.].
\bibitem[()]{Pines}D. Pines, \textsl{Elementary Excitations in Solids} (W. A. Benjamin, Inc., New York, 1963).
\bibitem[()]{Aoki}M. Aoki and H. Amawashi,
\href{http://arxiv.org/PS_cache/cond-mat/pdf/0702/0702257v1.pdf}{Cond-mat/0702257.}
\bibitem[()]{Keldysh1}L. V. Keldysh, \href{http://jetpletters.ac.ru/ps/426/article_6696.pdf}{Pisma v ZhETF {\bf 29}, 716 (1979)} [JETP Lett.]
\bibitem[()]{Iyengar} A. Iyengar, J. Wang, H. A. Fertig, and L. Brey,
\href{http://link.aip.org/getpdf/servlet/GetPDFServlet?filetype=pdf&id=PRBMDO000075000012125430000001}{Phys.
Rev. {\bf B 75}, 125430 (2007)}.
\bibitem[()]{Dharma}M. W. C. Dharma-wardana,
\href{http://link.aip.org/getpdf/servlet/GetPDFServlet?filetype=pdf&id=PRBMDO000075000007075427000001}
{Phys. Rev. {\bf B 75}, 075427 (2007).}
\end{thebibliography}
\end{document}